\documentclass[twocolumn,prl,amsmath,amssymb,showpacs]{revtex4-1}
\usepackage{epsf}      
\usepackage{graphicx}
\usepackage{color}
\usepackage{gensymb}

\begin{document}

\title {Anomalous magnetic and spin glass behavior in Nb substituted LaCo$_{1-x}$Nb$_x$O$_3$}
\author{Rishabh Shukla}
\author{R. S. Dhaka}
\email{rsdhaka@physics.iitd.ac.in}
\affiliation{Department of Physics, Indian Institute of Technology Delhi, Hauz Khas, New Delhi-110016, India}

\date{\today}                                         

\begin{abstract}
We report the structural, magnetic, transport and electronic properties of Nb substituted LaCo$_{1-x}$Nb$_x$O$_3$ ($x =$ 0--0.2). The Rietveld analysis of x-ray diffraction data demonstrate structural phase transitions from rhombohedral to orthorhombic and further to monoclinic with increasing the Nb concentration up to $x \ge$ 0.2. Interestingly, we observed dramatic changes in the magnetization (M) with increasing the Nb concentration, as the M sharply increases below 10~K even at 2.5\% substitution. Furthermore, ac susceptibility data show the spin-glass behavior in $x=$ 0.1 sample. We find that the density of states near the Fermi level decreases and the activation energy increases, which results in the decreasing conductivity with higher Nb concentration. A significant shift in the peak position of A$_{2g}$ phonon mode has been observed using Raman spectroscopy, which indicates the change in the coupling due to the structural distortion with Nb substitution. The core-level photoemission study confirms that the Nb is present in 5+ valence state. Our study reveals that the nonmagnetic Nb$^{5+}$ ($d^0$) substitution converts Co$^{3+}$ ions to Co$^{2+}$ and stabilize both in the high-spin state. Our results suggest that structural and spin-state transitions as well as the difference in the ionic radii between Nb$^{5+}$ and Co$^{3+}$ are playing an important role in tuning the physical properties. \\

PACS: 75.60.Ej, 75.60.Tt, 75.75.Cd, 75.70.Rf, 79.60.--i 

\end{abstract}

\maketitle

\section{\noindent ~Introduction}

The rare-earth cobaltite LaCoO$_3$ exhibit unusual magnetic and electronic phase transitions, which can be tuned by changing the temperature, pressure \cite{RaccahPRB67, VogtPRB03, LengsdorfPRB04, VankoPRB06, LengsdorfPRB07, KozlenkoPRB07} as well as with chemical substitutions at the La/Co site \cite{VasquezPRB96,KrienerPRB04,HammerPRB04,BaierPRB05,KreinerPRB09, SatoJPSJ14, StreltsovJPCM16, KarpinskyJPCM16, SerranoJPDAP08, YoshiPRB03}. The 3$d$ orbitals of the Co$^{3+}$ ion, surrounded with octahedrally coordinated O$^{2-}$ ions, are split into the t$_{2g}$ (triply degenerate) and the e$_g$ (doubly degenerate) orbitals in the lower and upper energy levels, respectively. The low temperature ground state is nonmagnetic (S=0) insulating nature with Co$^{3+}$ ions being in the low spin (LS) state configuration (3d$^6$; t$_{2g}^6$e$_g^0$) \cite{ZhuangPRB98}. However, with increasing temperature, the magnetic susceptibility ($\chi$) shows two prominent features/transitions at around 90~K and 500~K \cite{ZobelPRB02,PodlesnyakPRB06,EnglishPRB02}. The nonmagnetic to paramagnetic transition at 90~K is believed to be due to the change in the spin state of Co ions from LS to high spin state [HS; t$_{2g}^4$e$_g^2$ (S=2)] or intermediate spin state [IS; t$_{2g}^5$e$_g^1$ (S=1)] or both mixed in a certain ratio \cite{EnglishPRB02, KorotinPRB96, RaccahJAP68, IshikawaPRL04, YamaguchiPRB97, SaitohPRB97, HaverkortPRL06, KliePRL07}. The multiple spin states of Co ions (i.e. the change in the relative populations of t$_{2g}$ and e$_g$ levels) are due to the competition between the crystal field splitting and the intra-atomic Hund's exchange energy \cite{RaccahPRB67,SenarisJSSC95,CaciuffoPRB99,AsaiPRB89}. Note that the energy difference between these energies is rather small and the crystal field splitting energy strongly depends on the Co-O bond length, which leads to the unusual magnetic behavior in doped LaCoO$_3$ \cite{FitaPRB05}. The feature near 500~K is ascribed to an insulator to metal transition \cite{TokuraPRB98}, where the electrical conductivity increases by two orders of magnitude. Interestingly, these transitions are strongly affected by substituting alkali-earth elements at La site and transition metal elements at Co site \cite{VasquezPRB96,KrienerPRB04,HammerPRB04,BaierPRB05,KreinerPRB09}. For example, a well established case is the observation of a spin-cluster-glass insulator to ferromagnetic metal transition at about 20\% concentration of Sr in bulk La$_{1-x}$Sr$_x$CoO$_3$ \cite{WuPRB03}. Recently, we reported that Sr/Ca substitution (i.e. hole doping) in LaCoO$_3$ nanoparticles establish the IS/LS states of Co$^{3+}$/Co$^{4+}$ and observed significant enhancement in the magnetization \cite{Ravi17}. Here, the double exchange interactions between Co$^{3+}$ and Co$^{4+}$ are playing an important role in controlling the magnetic/spin-state transitions. Also, the layered cobaltites La$_{2-x}$Sr$_x$CoO$_4$ (containing both Co$^{2+}$ and Co$^{3+}$) are very interesting to understand the role of different spin as well as valence states of Co ions in controlling the magnetic, transport and electronic properties \cite{Hollmann, WuPRB10, ChangPRL09, CwikPRL09, BabkevichPRB10, TealdiPRB10, LiSR16}. For example, it has been reported that the Co$^{2+}$ present mostly in HS state, whereas different spin-states are possible in Co$^{3+}$ \cite{ChangPRL09, CwikPRL09}. The magnetic susceptibility behavior also indicate the presence of the Co$^{2+}$ ions in HS state and a possibility of magnetic ordering between Co$^{2+}$ and Co$^{3+}$ ions \cite{Hollmann}.     

In this context, the substitution of 3$d$ elements at the Co site in LaCoO$_3$ affects the spin state of nearby Co$^{3+}$ and exhibit various anomalous behaviors in the magnetic and transport properties \cite{TomiyasuPRB13, HammerPRB04, ViswanathanPRB09, Vinod14,Vinod13, RajeevanJMMM15, BullPRB16}. In LaCo$_{1-x}$Ni$_{1-y}$O$_3$, where the end members are a insulator (LaCoO$_3$) and a paramagnetic metal (LaNiO$_3$), a metal to insulator transition was observed at about 40\% Ni substitution \cite{HammerPRB04,RajeevPRB92}. Moreover, magnetization and neutron diffraction studies observed the presence of spin glass and long-range ordered ferromagnetic correlation at low temperatures \cite{HammerPRB04, ViswanathanPRB09, Vinod14, Vinod13, RajeevanJMMM15}. However, a recent study on highly metelllic LaNiO$_3$ single crystals by Li {\it et al.} revealed antiferromagnetic ordering in magnetization, specific heat and neutron scattering experiments \cite{LiNC18}, which is in contrast with another report in \cite{ZhangCGD}. In case of LaCo$_{1-x}$Mn$_x$O$_3$, where LaMnO$_3$ is antiferromagnetic insulator and crystallizes in an orthorhombic structure, detailed magnetization and neutron diffraction studies suggest the ferromagnetic/glassy nature and structural transition with Mn concentration \cite{BullPRB16}. Another interesting case is the substitution of 4$d$ elements like nonmagnetic Rh$^{3+}$, which is isoelectronic to Co$^{3+}$ and therefore, no charge transfer is expected \cite{KnizekPRB12, AsaiJPSJ, GuoPRB16}. Also, the 4$d$ orbitals are much more extended than 3$d$ orbitals, which results in the decreasing the correlation. It is reported that Rh$^{3+}$ substitution stabilizes the HS state of Co$^{3+}$ in LaCo$_{1-x}$Rh$_x$O$_3$ where the driving forces are the elastic and electronic interactions associated with the larger ionic radii and the unfilled 4$d$ shell of Rh$^{3+}$ cation \cite{KnizekPRB12}. Asai {\it et al.}, observed the ferromagnetic ordering below 15~K in the range of 10 to 40\% Rh concentration \cite{AsaiJPSJ}. The authors suggested that in this case the magnetic ordering is driven only by Co$^{3+}$ ions \cite{AsaiJPSJ}. This is different from the metallic ferromagnetic state of La$_{1-x}$Sr$_x$CoO$_3$, which is due to the mixed valence of Co$^{3+}$ and Co$^{4+}$ ions \cite{WuPRB03}. On the other hand, Guo {\it et al.}, did not observe signature of long range magnetic ordering till 4~K, but suggest a spin glass ground state in LaCo$_{1-x}$Rh$_x$O$_3$ \cite{GuoPRB16}.   

 Despite many theoretical and experimental investigations, the discussions on the nature of magnetic/spin-state transitions with chemical substitutions and their origin are still controversial. In fact Motohashi {\it et al.} observed both the ferromagnetic and spin glass states and the competition between them in SrCo$_{1-x}$Nb$_x$O$_{3-\delta}$ \cite{MotohashiPRB05, MotohashiAPL05}. Therefore, it is vital to further investigate the physical properties with substituting 4$d$ elements, which introduces itinerant electrons into the system. Recently, Oygarden {\it et al.}, reported the structural transition from rhombohedral to orthorhombic/monoclinic and reduction in the electrical conductivity with Nb substitution in LaCo$_{1-x}$Nb$_x$O$_3$, which are discussed in terms of spin state of Co ion \cite{Oygarden}. They show that as the valence state of Co ion is 3+ in LaCoO$_3$, the charge balance with Nb substitution is by converting Co$^{3+}$ to Co$^{2+}$ where the chemical formula can be expressed as LaCo$^{3+}_{1-3x}$Co$^{2+}_{2x}$Nb$^{5+}_x$O$_3$ i.e. Co is completely reduced to Co$^{2+}$ for $x = $0.33 sample \cite{Oygarden}. Also, the ionic radius of Nb$^{5+}$ (0.64~\AA~)/Co$^{2+}$ [0.65~\AA~ in LS (S = 0.5) and 0.745~\AA~ in HS (S = 1.5)] is larger than that of the Co$^{3+}$ (0.545~\AA~ in LS, 0.56~\AA~ in IS and 0.61~\AA~ in HS) ions, which results increasing the average radius at the Co position and induce the lattice expansion \cite{Oygarden}. Interestingly, the simultaneous substitution of Sr and Nb in LaCoO$_3$ prevent the creation of Co$^{4+}$ and it is reported that exchange interactions between IS/HS Co$^{3+}$ can induce ferromagnetism \cite{SikolenkoJPCM09, Troyanchuk}.

In order to understand the role of Co valence state for achieving the ferromagnetism, spin glass and spin-state transition in LaCoO$_3$, we study the structural, magnetic, transport and electronic properties of nonmagnetic Nb$^{5+}$ ($4d^0$) substituted LaCo$_{1-x}$Nb$_x$O$_3$. The Rietveld refinements of XRD data clearly indicate the structural phase transition for $x \ge$ 0.1 samples. More interestingly, we observed dramatic changes in the magnetization for Nb concentration as low as 2.5\%. Furthermore, the spin glass behavior has been observed for $x=$ 0.1 sample. Our resistivity measurements show the strong insulating nature with increasing Nb concentration. The XPS study confirms that the Nb is present in 5+ valence state. We find that nonmagnetic Nb$^{5+}$ substitution converts Co$^{3+}$ ions to Co$^{2+}$ and stabilize in the high-spin state. We discuss the possible role of spin-state transition and the difference in the ionic radii between Nb$^{5+}$ and Co$^{3+}$/Co$^{2+}$. 

\section{\noindent ~Experimental Details}

The polycrystalline samples of LaCo$_{1-x}$Nb$_x$O$_3$ were synthesized by the conventional solid state reaction method. The starting materials Co$_3$O$_4$ (99.99\% Alfa) and Nb$_2$O$_5$ (99.9\% Sigma) were used as received, without further purification. But, the La$_2$O$_3$ (99.9\% Sigma) powder was dried prior to use at 700$^0$C for 12 hrs due to its hygroscopic nature. The stoichiometric amount of starting materials were thoroughly mixed with the help of mortar-pestle and reacted at 900$^0$C for 8 hour in air, calcined powder was pressed into pellets at 2000 psi with a hydraulic press and sintered in air at 1200$^0$C for the 24 hrs, in case of substituted samples temperature profile was 1100$^0$C for 12 hrs followed by 1475$^0$C for 6 hrs \cite{Oygarden}.

The powder x-ray diffraction (XRD) data were collected with CuK$\alpha$ radiation (1.5406 \AA ) from Panalytical X-ray diffractometer, we analyzed the XRD data by Rietveld refinement using FullProf package and the background was fitted using linear interpolation between the data points. The magnetic and transport measurements were performed with physical property measurement system (PPMS) from Quantum design, USA. A commercial electron energy analyzer (PHOIBOS 150 from Specs GmbH, Germany) and a non-monochromatic AlK$\alpha$ x-ray source (h$\nu$ = 1486.6~eV of line width $\sim$0.9~eV) used for XPS measurements with the base pressure in the range of 10$^{-10}$ mbar. We analyzed the core-level spectra after subtracting an inelastic Tougaard background. AC susceptibility measurements were performed with Magnetic property measurement system (MPMS) from Quantum design, USA. The Raman measurements were carried out in backscattering geometry using a Renishaw inviaconfocal Raman microscope using unpolarized laser of 514~nm and 785~nm excitation wavelength, 2400 lines per mm grating and 0.2~mW laser power at room temperature.

\section{\noindent ~Results and Discussion}

\begin{figure}
\includegraphics[width=3.5in]{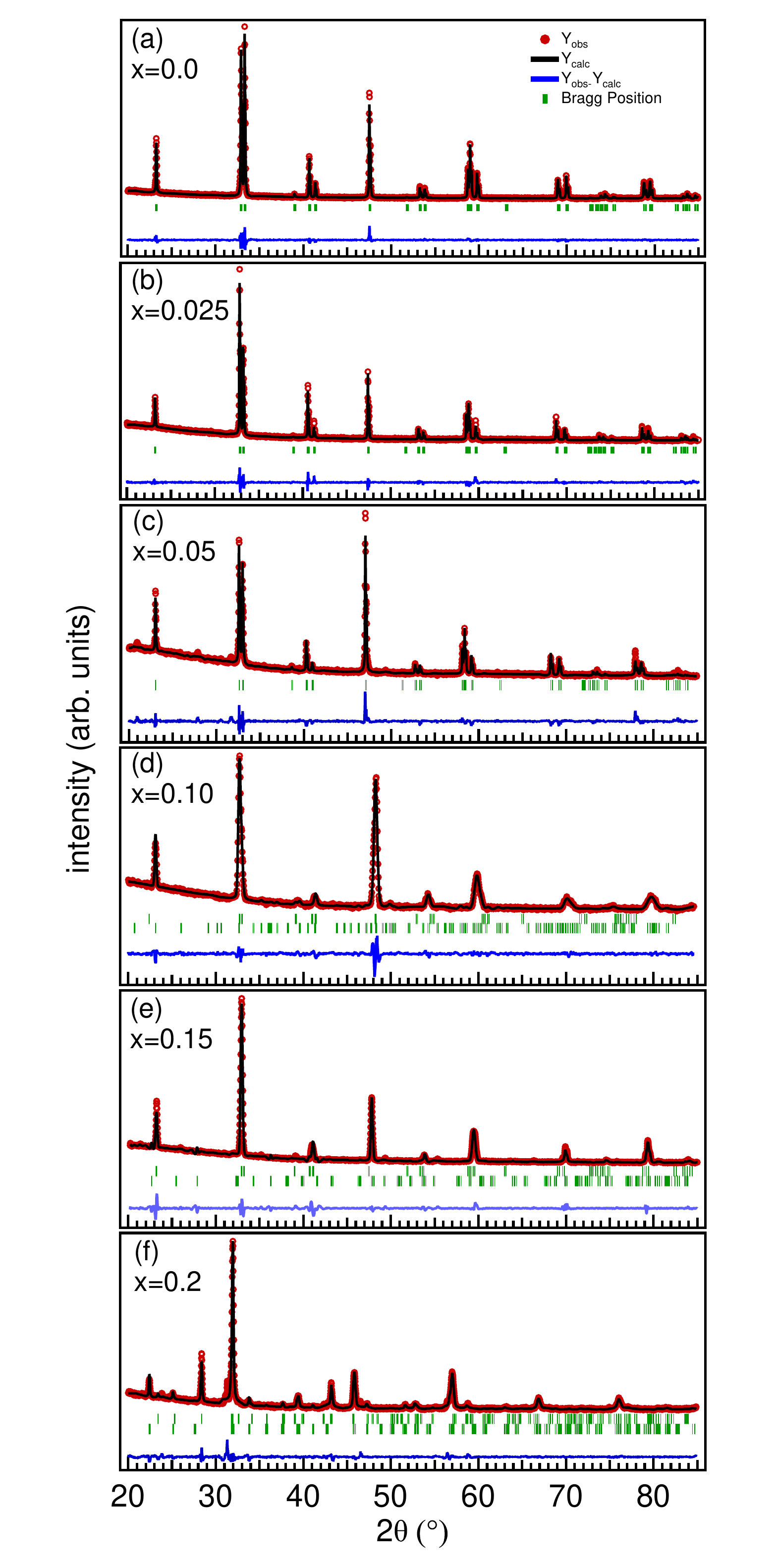}
\caption{The room temperature powder x-ray diffraction data (red circles) of LaCo$_{1-x}$Nb$_x$O$_3$ ($x=$0 -- 0.2). The Rietveld refinement profiles, the fitted Bragg peak positions, and the residual are shown in black solid lines, short vertical bars and blue lines in the bottom of each panel, respectively.} 
\label{fig1:xrd}
\end{figure}

In Figs.~1(a--f), we show the room temperature powder x-ray diffraction data of LaCo$_{1-x}$Nb$_x$O$_3$ ($x=$ 0 -- 0.2) with increasing Nb concentration. The Rietveld refinements confirm the rhombohedral structure (space group R$\bar{3}$c) for $x=$ 0, 0.025 and 0.05 samples, as shown in panels (a, b, c) of Fig.~1. We set the background with linear polynomial during the Rietveld refinements. With further increasing Nb concentration, the XRD patterns could not be fitted with only rhombohedral phase. Therefore, we have added the orthorhombic phase (space group Pbnm) for $x =$ 0.1 and 0.15 samples, which produce the fitting of the data well. Further, we have quantified the contribution of orthorhombic phase and observed the ratio of about R$\bar{3}$c:Pbnm = 74:26 and 58:42 in $x=$ 0.1 and 0.15 samples, respectively. With further increasing the Nb concentration, we observed a monoclinic distortion as the XRD pattern of $x=$ 0.2 sample can be fitted with orthorhombic and monoclinic phases in the ratio of Pbnm:P2$_1$/n = 62:38. Also, we observed a shift in the peak position (2$\theta$ $\approx$ 32.8$\degree$) towards the lower 2$\theta$ side with higher Nb content, which indicates the increment in the interplanar spacing as $d$ $\alpha$ 1/$\sin\theta$. Our analysis clearly reveals the first-order structural phase transition with Nb substitution, and the coexistence of two phases at $x\ge$0.1 is in agreement with ref.~\cite{Oygarden}. The possibility for this transformation can be thought as, due to the difference between the ionic radii of Nb$^{5+}$/Co$^{2+}$ and Co$^{3+}$. Thus due to larger ionic radii of Nb$^{5+}$/Co$^{2+}$, the Goldschmidt tolerance factor {\it t} decreases with Nb substitution and the volume of octahedra around B site increases, which results in the structural phase transition from rhombohedral to orthorhombic and then to monoclinic with higher Nb concentration \cite{Oygarden}. This is also consistent with a recent study by Guo {\it et al.}, which report that the larger size Rh$^{3+}$ (0.665 \AA) substitution at Co cite, rapidly stabilizes the Pbnm (orthorhombic) structure \cite{GuoPRB16}. Furthermore, for $x=$0 sample, the calculated Co--O--Co bond-angle is found to be 162.1$^{\rm o}$, and six Co--O bonds, two La--Co bonds and other six La--Co bonds are with average length of 1.936~$\rm \AA$, 3.273~$\rm \AA$, and 3.326~$\rm \AA$, respectively. Also, there are six Co--Co bonds at average length 3.825~$\rm \AA$, these parameters are in good agreement with reported values \cite{ KnizekPRB05, PandeyJPCM06}. The pseudocubic (pc) lattice parameters, which are calculated using a$_{pc}$=a/$\sqrt{2}$, b$_{pc}$=b/$\sqrt{2}$, c$_{pc}$=c/$\sqrt{12}$ for R$\bar{3}$c, and c$_{pc}$=c/2 for Pbnm and P2$_1$/n, and unit cell volume are presented in table~I. We note that the bond-angle and cell volume are increasing up to $x=$0.05 and then the bond-angle decreases for $x=$ 0.1 and 0.15 samples where refinement has been performed with two space groups (rhombohedral and orthorhombic) \cite{MartiJPCM94}. The Rietveld analysis of all the samples show good quality of fitting and the refined parameters (including bond-angle and average bond length) of LaCo$_{1-x}$Nb$_x$O$_3$ ($x=$ 0 -- 0.2) are summarized in table~I.   

\begin{table*}
		\centering
		\label{tab:rietveld}
		\caption{The Rietveld refined parameters from the XRD data of LaCo$_{1-x}$Nb$_x$O$_3$ ($x=$ 0 -- 0.2) are summarized along with the pseudocubic (pc) lattice parameters and unit cell volume.}
		\vskip 0.2cm
		\begin{tabular}{|c|c|c|c|c|c|ccccc|c|c|c|c|}
		\hline
		\textbf{$x$}&\textbf{$\chi^2$}&\textbf{a (\normalfont\AA)} &\textbf{b (\normalfont\AA)}&\textbf{c (\normalfont\AA)} & space group & Co-O(\AA) & Co--O--Co & La-Co(\AA) & La-Co(\AA)& Co-Co(\AA) &\textbf{a$_{\rm pc}$} &\textbf{b$_{\rm pc}$}&\textbf{c$_{\rm pc}$} & \textbf{V$_{\rm pc}$} \\
		
		&& &&& (wt$\%$)  & $\times$6 & & $\times$2 & $\times$6 & $\times$6 &(\normalfont\AA)  &(\normalfont\AA)&(\normalfont\AA) &(\normalfont\AA$^3$)  \\
		\hline
		0 &1.75&5.442&5.442&13.090& R$\bar{3}$c--100\% & 1.936 & 162.1& 3.273 & 3.326 & 3.825&3.848 &3.848&3.779& 56.0 \\
		\hline
		0.025 &1.72&5.454&5.454&13.117& R$\bar{3}$c--100\% & 1.921 & 172.4 & 3.279 & 3.333& 3.833&3.857&3.857&3.885& 56.3 \\
		\hline
		0.05 &2.32&5.493&5.493&13.214&R$\bar{3}$c--100\% & 1.934 & 173.8 & 3.304 & 3.357& 3.861&3.885&3.885&3.815 &57.6 \\
		\hline
		0.10 &2.2&5.132&5.132&12.439&R$\bar{3}$c--74\% & 1.816 & 169.6 &3.11 & 3.139& 3.616&3.630&3.630&3.59&47.3 \\
		&&5.451&5.698&7.747&Pbnm--26\% &-- & --&-- & --&--&3.855&4.029&3.874 &60.2 \\
		\hline
		0.15 &2.85&5.436&5.436&13.172&R$\bar{3}$c--58\% & 1.927 & 157.2 & 3.216 & 3.289& 3.777&3.845&3.845&3.803 &56.3 \\
		&&5.543&5.504&7.845& Pbnm--42\% &-- & --&-- & --&--&3.920&3.893&3.923& 59.9 \\
		\hline
		0.20 &2.35&5.604&5.633&7.599&Pbnm--62\% &-- & --&-- & --&--&3.963&3.984&3.799 &60.0 \\
		&&5.589&5.626&7.909& P2$_1$/n--38\% &-- & --&-- & --&-- &3.953&3.978&3.955& 55.5 \\
		\hline
		\end{tabular}
\end{table*}

It is interesting to investigate the magnetic behavior with the substitution of nonmagnetic Nb$^{5+}$ ($d^0$) at Co$^{3+}$ site in LaCoO$_3$. Therefore, we present in Figs.~2 (a--f) the isothermal magnetization data of LaCo$_{1-x}$Nb$_x$O$_3$ ($x=$ 0 to 0.2) measured at 5~K. The samples are first cooled to 5~K in zero field and then M--H loops are obtained as a function of applied magnetic field. For the parent $x=$ 0 sample, the value of coercivity and spontaneous magnetization (M$_{\rm S}$) are observed about 1.35~kOe and 105~emu/mol [Fig.~2(a)], which indicate the presence of weak ferromagnetism. At the same time, the non-saturation nature of the magnetization up to 70~kOe [Fig.~2(a)] suggest the canted antiferromagnetism in the sample \cite{Vinod14}, which could be due to the superexchange interaction between Co$^{3+}$--Co$^{3+}$ ions \cite{SeoPRB12, MerzPRB10, ZhangPRB12}. Interestingly, the M$_{\rm S}$ value is significantly higher (600~emu/mole) even with 2.5\% Nb substitution, which further increased with increasing Nb concentration up to 10\% (i.e. M$_{\rm S}=$ 1500 and 2000~emu/mole for $x=$ 0.05 and 0.1 samples, respectively). With increasing the Nb concentration, the M$_{\rm S}$ value start decreasing and for $x=$ 0.2 sample we observed a paramagnetic behavior. Moreover, the small hystersis present in $x=$ 0 sample disappeared completely for $x\ge$ 0.025 samples. It is interesting to note that the M-H data for 0.025 $\le x \le $ 0.15 samples [see Figs.~2(b--e)] exhibit a superparamagnetic like nature, having negligible values of coercive field and remanance, but very high value of the magnetization. However, for $x > $ 0.15 a paramagnetic behavior starts dominating, for example almost a straight line behavior has been observed for $x=$ 0.2 sample, see Fig.~2(f). As the Nb$^{5+}$ has d$^0$ configuration, it is thought to be act as diamagnetic dilution in the solid solution of LaCo$_{1-x}$Nb$_x$O$_3$, which decreases the possibility of any long-range magnetic ordering. The Nb$^{5+}$ substitution increases the population of Co$^{2+}$ ions via changing the valency of nearest neighbor Co$^{3+}$ ions, which play an important role in tuning the structural and magnetic phase transitions in LaCoO$_3$.   

\begin{figure}
\centering
\includegraphics[width=3.6in]{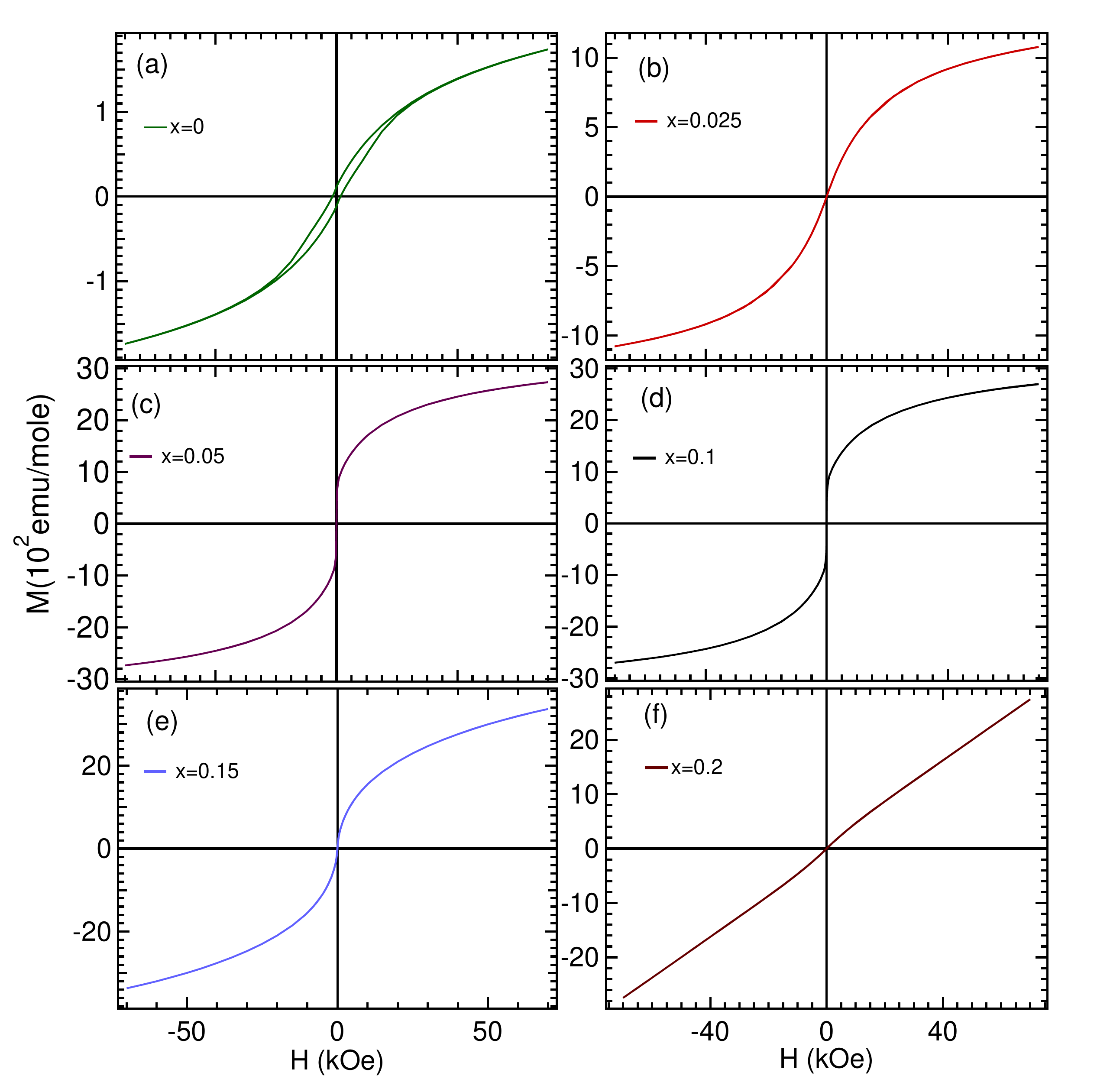}
\caption {Isothermal magnetization data of LaCo$_{1-x}$Nb$_x$O$_3$, $x=$ 0 to 0.2, recorded at 5~K sample temperature.}
\label{fig2:MH}
\end{figure} 

\begin{figure}
\centering
\includegraphics[width=3.5in]{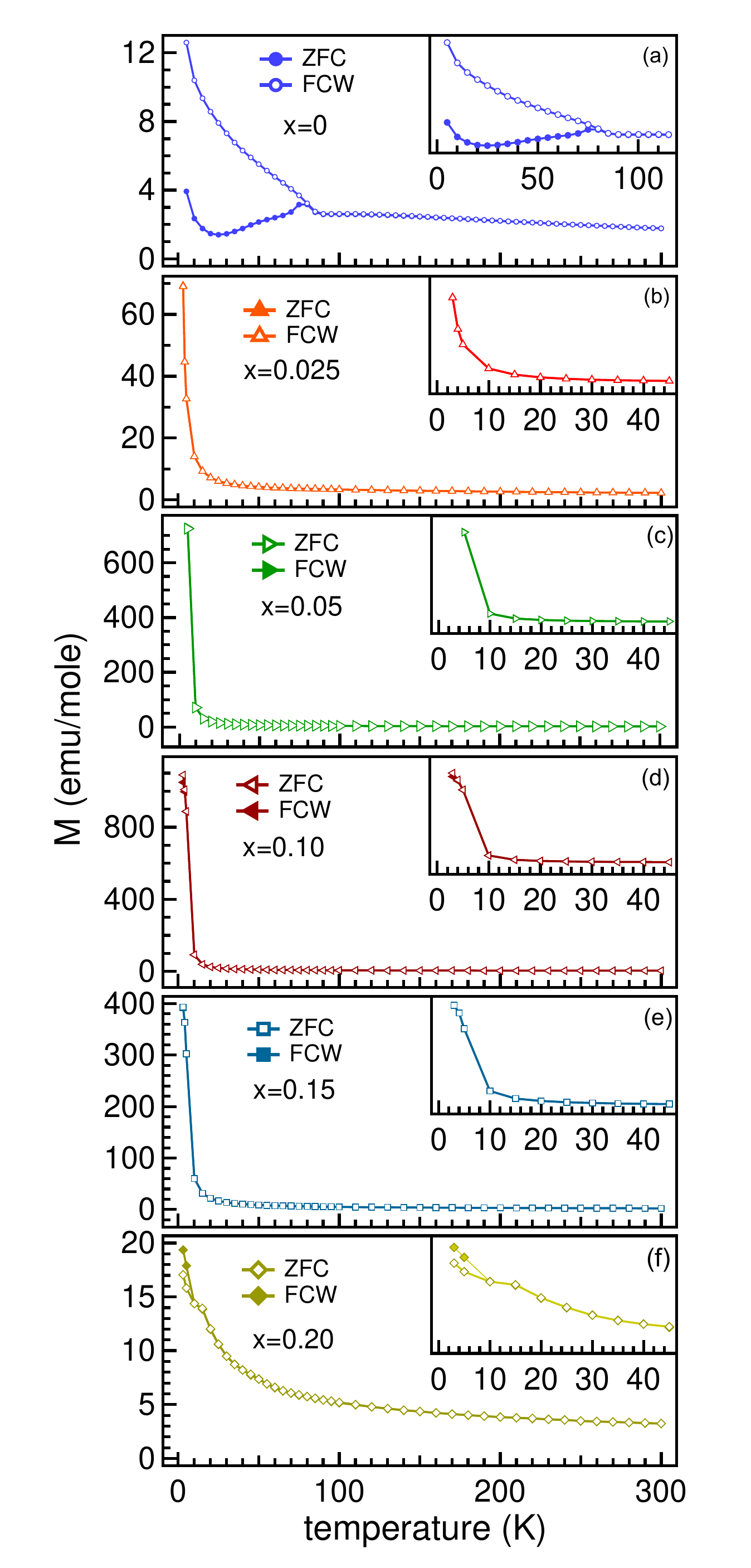}
\caption {The magnetization (ZFC \& FCW) of LaCo$_{1-x}$Nb$_x$O$_3$, $x=$ 0--0.2 (a--f) as a function of temperature measured at applied magnetic field of 500~Oe, insets in each panel show the illustrated view in the low temperature region.}
\label{fig3:MT}
\end{figure}

\begin{figure}
\centering
\includegraphics[width=3.6in]{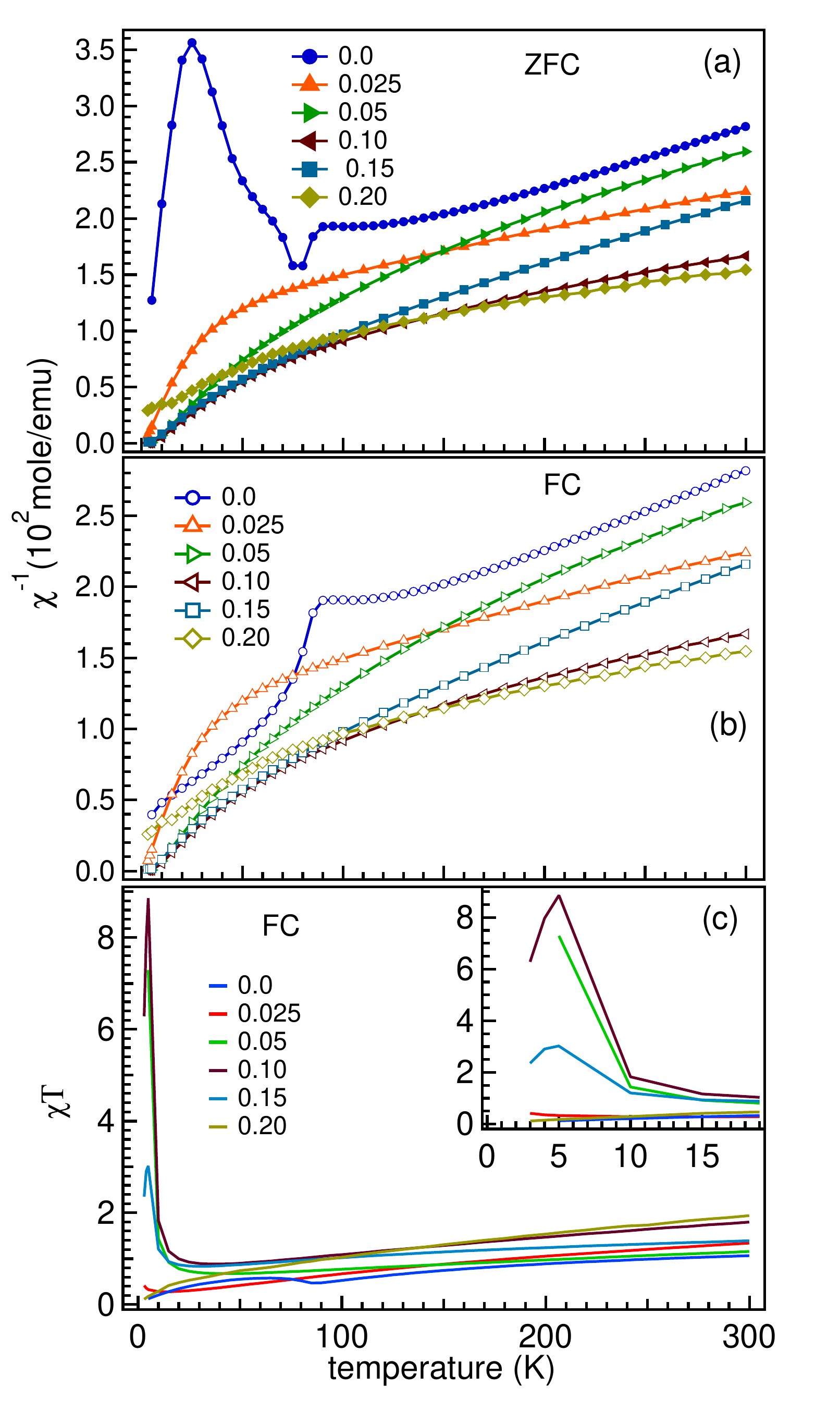}
\caption {Inverse susceptibility ($\chi^{-1}$) of LaCo$_{1-x}$Nb$_x$O$_3$, $x=$ 0 to 0.2 as a function of temperature measured at H = 500~Oe both in ZFC (a) and FC (b) modes. (c) the plot between $\chi$T vs temperature (FC) of all the samples. Inset in (c) is the zoomed view at low temperature below 20~K.}
\label{fig4:Chi-1}
\end{figure}

We therefore study the temperature dependence of the magnetization behavior of LaCo$_{1-x}$Nb$_x$O$_3$, $x=$ 0 to 0.2, as measured at applied magnetic field of 500~Oe and presented in Fig.~3. For $x=$ 0 sample, a clear ferromagnetic transition at T$_{\rm C}$ $\approx$80~K has been observed in the zero field cooled -- field cooled (ZFC--FC) data \cite{Vinod14}, and below T$_{\rm C}$ the ZFC magnetization decreases from 2.5~emu/mole to 1.5~emu/mole till about 20~K, and then increases back to about 4~emu/mole at 5~K. The FC magnetization increases slowly till 20~K and then there is a fast increase up to 18~emu/mole at 5~K. A large bifurcation in ZFC and FC curves below 80~K indicates the possibility of short-range ferromagnetic interactions. Above T$_{\rm C}$, the ZFC-FC data show a paramagnetic behavior, where the magnetic susceptibility obeys the Curie-Weiss law (fitted in 200 to 300~K range). Note that only 2.5\% Nb substitution changes the ZFC-FC behavior drastically, as can be seen in Figs.~3(b--f). For $x=$ 0.025 to 0.15 samples, there is no significant change in the magnetization till about 10~K and below this there is a sharp increase in the magnetization upto the lowest measured temperature. The magnetization values at 5~K increases from 70~emu/mole for $x=$ 0.025 sample to 700~emu/mole and 900~emu/mole for $x=$ 0.05 and 0.1 samples, respectively. Further increasing Nb concentration, it decreases to about 400~emu/mole for $x=$ 0.15 sample. However, there is no bifurcation between FC and ZFC magnetization data for $x=$ 0.025 -- 0.15 samples. The ZFC-FC results of $x=$ 0.025 -- 0.15 samples indicate the formation of small clusters with the spins aligned in ferromagnetic order and these spins freezes at the blocking temperature of $\le$10~K. Interestingly, a continuous increase in the magnetizaion has been observed for $x=$ 0.2 sample [Fig.~3(f)], with a slight difference in the FC and ZFC data below $\approx$10~K. Note that for $x=$ 0.2 sample, the magnetization is significantly lower i.e. about 20~emu/mole at 5~K. 

\begin{table}
  \centering
  \caption{Experimentally obtained and calculated parameters, Curie constant (C) in emu K mol$^{-1}$, effective magnetic moment ($\mu_{eff}$) in $\mu_B$, of LaCo$_{1-x}$Nb$_x$O$_3$.}
  \vskip 0.2cm
  \label{tab:table1}
  \begin{tabular}{|c|c|c|c|c|c|c|c|c|c|c|c|}
  	\hline
   \textit{x} & $\theta_{\rm CW} $ & C &$\mu_{eff}$  & S$_{avg}$&$\mu_{eff}$  & S$_{avg}$& (1-2$x$) & 2$x$\\
   
     & (K) &  & (exp) &  (exp)& (cal) &  (cal)&  $\times {\rm Co}^{3+}$ &  $\times {\rm Co}^{2+}$\\
     
     &  &  & &  & & & IS : HS &  HS\\
    \hline
    0      &-220 & 1.85 & 3.85 & 1.5 & 3.85 & 1.5 & 50 : 50 & 0\\
    \hline
    0.025     &-329 & 2.8 & 4.70 & 1.9 &4.65 & 1.88& 10 : 90  & 100 \\
    \hline
    0.10      &-206 & 3.0  & 4.90 & 2.0  &4.7  & 1.9& 0 : 100 &100\\
    \hline
    0.20   &-289 & 3.75 & 5.50 & 2.3 &4.5 &1.8 & 0 : 100 &100 \\
    \hline
  \end{tabular}
\end{table}

Furthermore, we analyze the inverse susceptibility ($\chi^{-1}$) vs temperature data, as shown in Figs.~4(a, b), from 200 to 300~K by the Curie-Weiss law 
\begin{equation}
\chi^{-1} = \frac{H}{M} = \frac{3k_B}{{\mu^2_{eff}}}(T-\theta_{\rm CW})
\end{equation}
and evaluate the effective magnetic moment $\mu_{eff}$ and the Curie-Weiss temperature $\theta_{\rm CW}$. The obtained values of $\mu_{eff}$ and $\theta_{\rm CW}$ for $x=$ 0 sample are consistent with the reported values in ref.~\cite{Vinod14}. The parameters obtained for all the samples with Nb substitution are summerized in table~II. As the Nb$^{5+}$ substitution at Co$^{3+}$ site will convert the neighboring Co ions from 3+ to 2+ valence states, it is important to examine the spin state of Co ions with Nb substitution and their role in controlling the structural and magnetic properties of LaCo$_{1-x}$Nb$_x$O$_3$. Here we use the approximation of spin only (as angular momentum is quenched) magnetic moment, S$_{av}$ and calculate the values, using experimentally obtained values of $\mu_{eff}$ with the formula $ \mu_{eff} = 2\sqrt{S(S+1}$, which are given in the table~II. For $x=$ 0 sample, the spin state contribution can be calculated by considering the mixed states of IS:HS in 50:50 ratio or LS:HS in 25:75 ratio, which gives the S$_{av}$ and $\mu_{eff}$ values very close to the experimental values. As it is reported in ref.~\cite{Oygarden} and also mentioned above in the Introduction section, at about 33\% Nb substitution Co$^{3+}$ ions will be completely converted to Co$^{2+}$, i.e. at this concentration there will be only Co$^{2+}$ ions present in the system. This means each Nb$^{5+}$ converts two Co$^{3+}$ ions to Co$^{2+}$ ions. We have considered this and calculated the total spin using S$_{avg} =$ (1$-2x$)Co$^{3+}$ + 2$x$Co$^{2+}$. Note that Nb$^{5+}$, Co$^{3+}$ (HS) and Co$^{2+}$ (LS) are similar in size and larger than Co$^{3+}$ (IS). We used different combinations of possible spin-states, and our calculations reveal that Nb substitution establish Co$^{3+}$ ions in the HS state up to 20\% concentration as the ionic radii of Nb$^{5+}$ and Co$^{3+}$ HS are very close. Interestingly, we find that Nb$^{5+}$ substitution converts two Co ions in to 2+ state and Co$^{2+}$ stabilize in high-spin state. We summarized the resulting values in table~II.  

\begin{figure}
\centering
\includegraphics[width=3.55in]{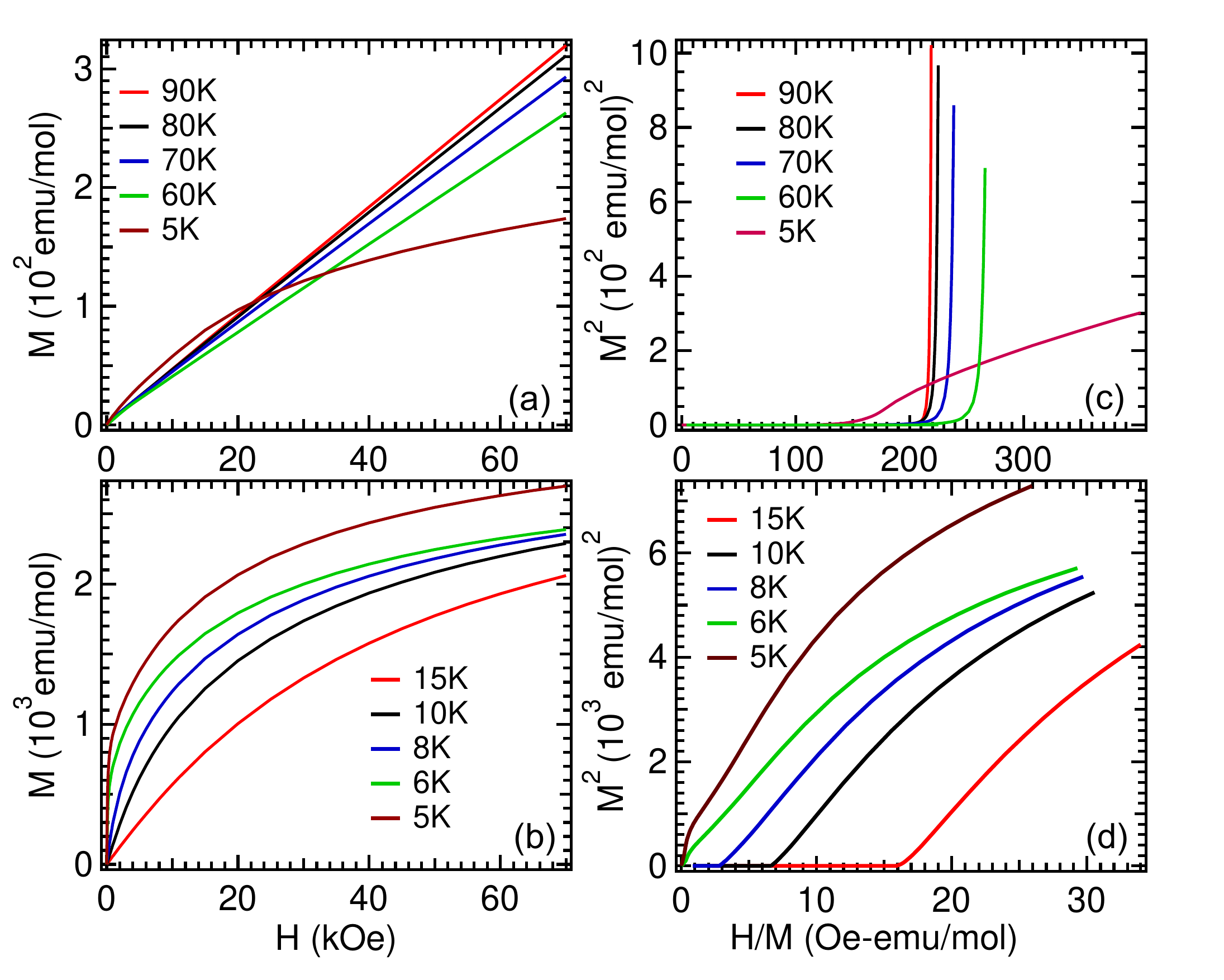}
\caption {(a, b) The virgin curves (M -- H) and (c, d) Arrott plots (M$^2$ vs H/M) of LaCo$_{1-x}$Nb$_x$O$_3$, $x=$ 0 and 0.1, respectively, at different temperatures across the transition.}
\label{fig5:Arrott}
\end{figure}  

As discussed above, for $x=$ 0 sample a clear transition is visible at $\approx$80~K and for $x=$ 0.025 to 0.15 samples the magnetization data show sharp increase below about 10~K. Therefore, we further examine the magnetic behavior of $x=$ 0 and 0.1 samples near their respective transition temperatures. Figs.~5(a, b) show the virgin curves (M -- H) measured at various temperatures across the transitions. For $x=$ 0 sample, a linear behavior is observed with slight decrease in the moment with lowering the temperature from 90~K to 60~K, see Fig.~5(a). On the other hand, the slope change in magnetization of $x=$ 0.1 sample is clearly seen in Fig.~5(b) where the magnetic moment at 70~kOe increases with lowering the temperature. We performed further analysis of the isothermal magnetization data [from Figs.~5(a, b)] and present Arrott's plot (M$^2$ vs H/M)\cite{Arrott} in Figs.~5(c, d). In the Arrott's plot, we fit the high field data by a straight line and extract the information by positive/negative intercept, which tells the ferro/non-ferro magnetic nature of the system. Also, the strength of spontaneous magnetization in the system can be estimated by the value of the intercept. For $x=$ 0 sample, see Fig.~5 (c), the intercept at all the temperatures is found to be negative, which suggest the presence of short-range magnetic ordering. However, for $x=$ 0.1 sample, the intercept upto 10~K is negative, which changes to positive at $\le$ 8~K, see Fig.~5(d). These results indicate the transition from short-range to long-range magnetic nature at low temperatures in $x=$ 0.1 sample. Though, the substitution of nonmagnetic Nb$^{5+}$ is expected to dilute the magnetic interactions, it creates Co$^{2+}$ in high-spin state and enhance the magnetic moment. 

\begin{figure}
\centering
\includegraphics[width=3.62in]{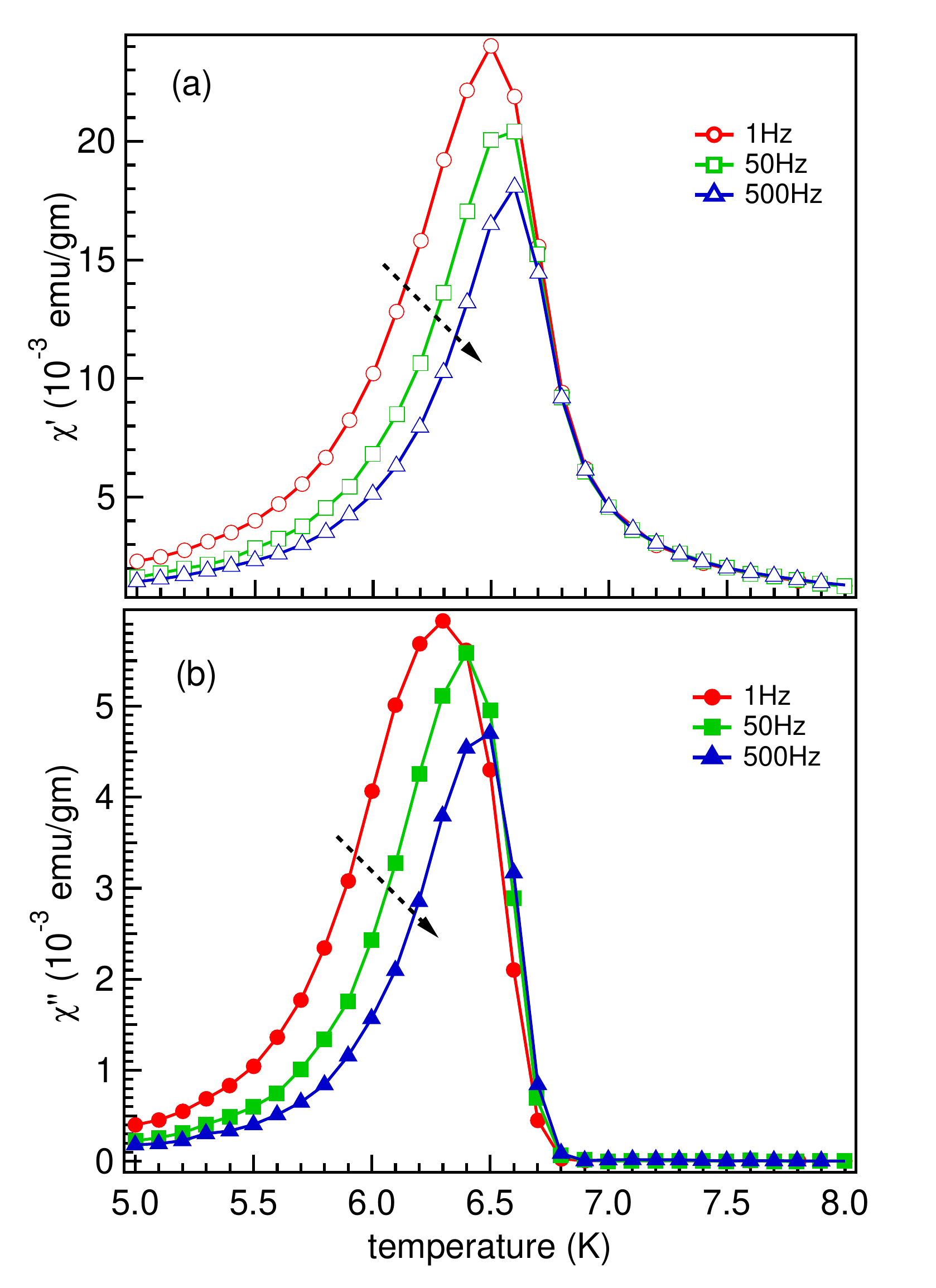}
\caption {ac-susceptibility data of LaCo$_{0.9}$Nb$_{0.1}$O$_3$ (a) real and (b) imaginary components, measured at H$_{ac}$ = 3.5~Oe and at different frequencies (1, 50 and 500~Hz).}
\label{fig6:AC}
\end{figure}

\begin{figure}
\centering
\includegraphics[width=3.62in]{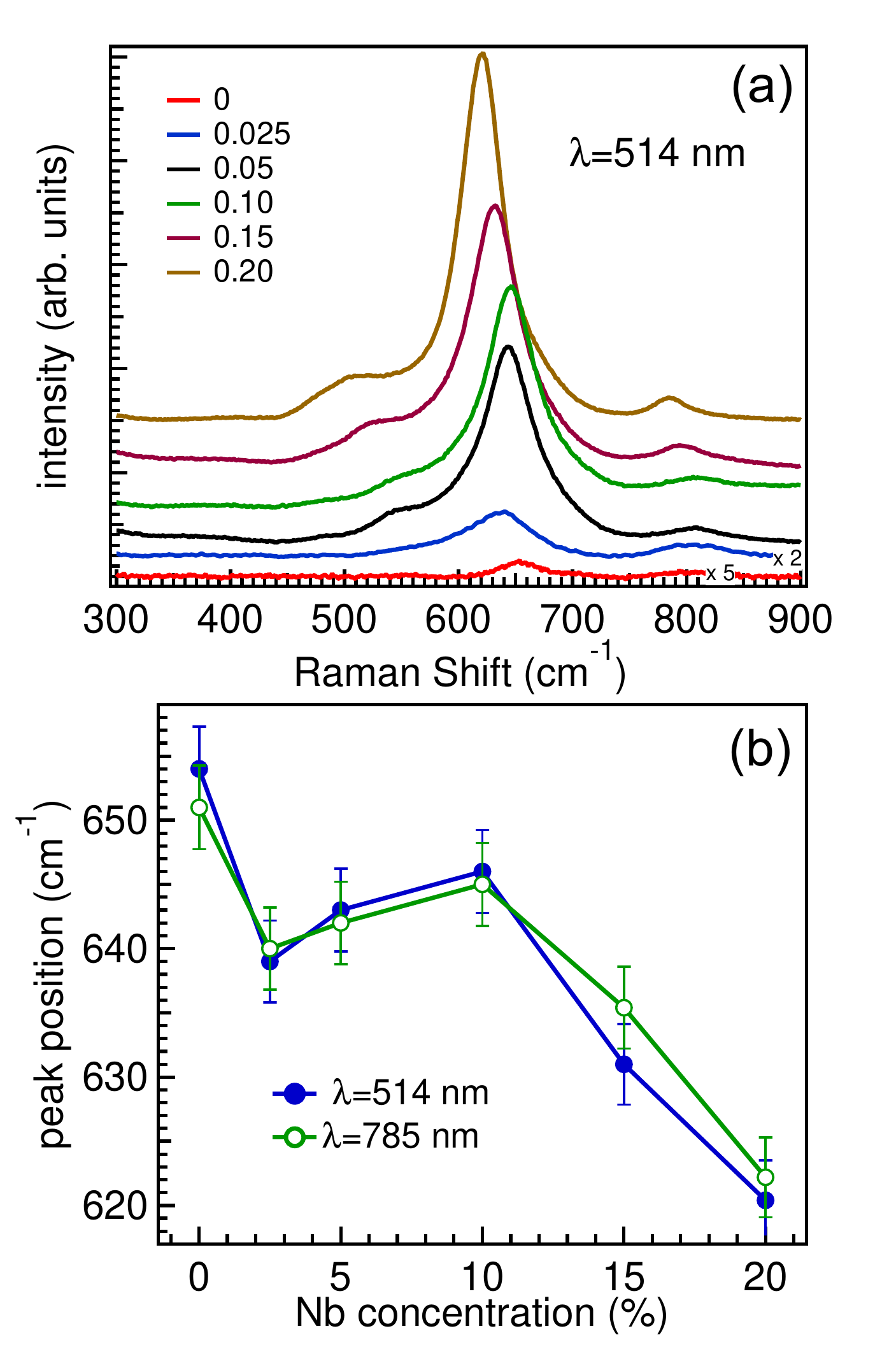}
\caption {(a) Raman spectra of LaCo$_{1-x}$Nb$_x$O$_3$ for different Nb concentration, measured with 514~nm wavelength. (b) Peak position of A$_{2g}$ mode as a function of Nb concentration, measured at room temperature using two excitation wavelengths.}
\label{fig7:Raman}
\end{figure}

In order to understand the presence of spin-glass/cluster spin-glass/superparamagnetic nature in these samples, the above analysis motivated to perform ac susceptibility measurements. Moreover, in Fig.~4(c), we observed a peak in the plot between $\chi$T vs temperature, at around 5~K for $x=$ 0.05 -- 0.15 samples where the peak is pronounced for $x=$ 0.1 sample. Therefore, to further examine, we performed the ac susceptibility measurements in the zero field cooled mode for LaCo$_{0.9}$Nb$_{0.1}$O$_3$ sample. Figs.~6(a, b) show both the real and imaginary components measured in the ac field H$_{ac}$ = 3.5~Oe at different frequencies (1, 50 and 500~Hz) in the temperature range from 5~K to 8~K. At 1~Hz, the real and imaginary parts show a peak at 6.5~K and 6.3~K, respectively. The peaks exhibit a clear shift to higher temperatures and change in the intensity with increasing the frequencies in both real and imaginary contributions. This behavior clearly indicate the presence of spin glass state in LaCo$_{0.9}$Nb$_{0.1}$O$_3$.      

Raman scattering measurements are useful to probe the coupling between the optical phonon modes and relate to the distortions of CoO$_6$ octahedra (tilting and rotation) with increasing Nb content. Also, the change in the peak position and relative intensity of the corresponding modes give evidence for the presence of different spin-states of Co ions. It has been reported that for the rhombohedral (R$\bar{3}$c) structure, group theoretical analysis gives five Raman active modes (A$_{1g}$ + 4E$_g$) \cite{KozlenkoPRB07,IshikawaPRL04, GranadoPRB98, AbrashevPRB99}. In Fig.~ 7(a), we show the room temperature Raman spectra of LaCo$_{1-x}$Nb$_x$O$_3$ ($x=$ 0 -- 0.2) in the range between 300 and 900 cm$^{-1}$. For $x=$ 0 and 0.025 samples, the intensity is very week, so we multiplied the intensity by a factor of 5 and 2, respectively, for clear presentation. We observed three peaks at about 560, 650 and 800 cm$^{-1}$, which can be assigned to the E$_g$ quadrupole, A$_{2g}$ breathing (stretching like internal vibrations of the CoO$_6$ octahedra), and another E$_g$ phonon modes, respectively \cite{KozlenkoPRB07,IshikawaPRL04, GranadoPRB98, AbrashevPRB99}. Interestingly, a clear shift has been observed in peak position of these modes. We now focus on the pronounced A$_{2g}$ mode and plotted the peak position measured using 514~nm and 785~nm excitation wavelengths, as shown in the Fig.~7(b). The shift in the A$_{2g}$ peak position indicates the change in coupling and the Co--O bond length i.e. further distortion in CoO$_6$ octahedra, which may favor the HS state of Co ions \cite{RadaelliPRB02}, consistent with the magnetic data analysis. Also, a significant increase in the peak intensity of A$_{2g}$ mode is evident [Fig.~7(a)] with increasing Nb concentration. 

\begin{figure}
\centering
\includegraphics[width=3.5in]{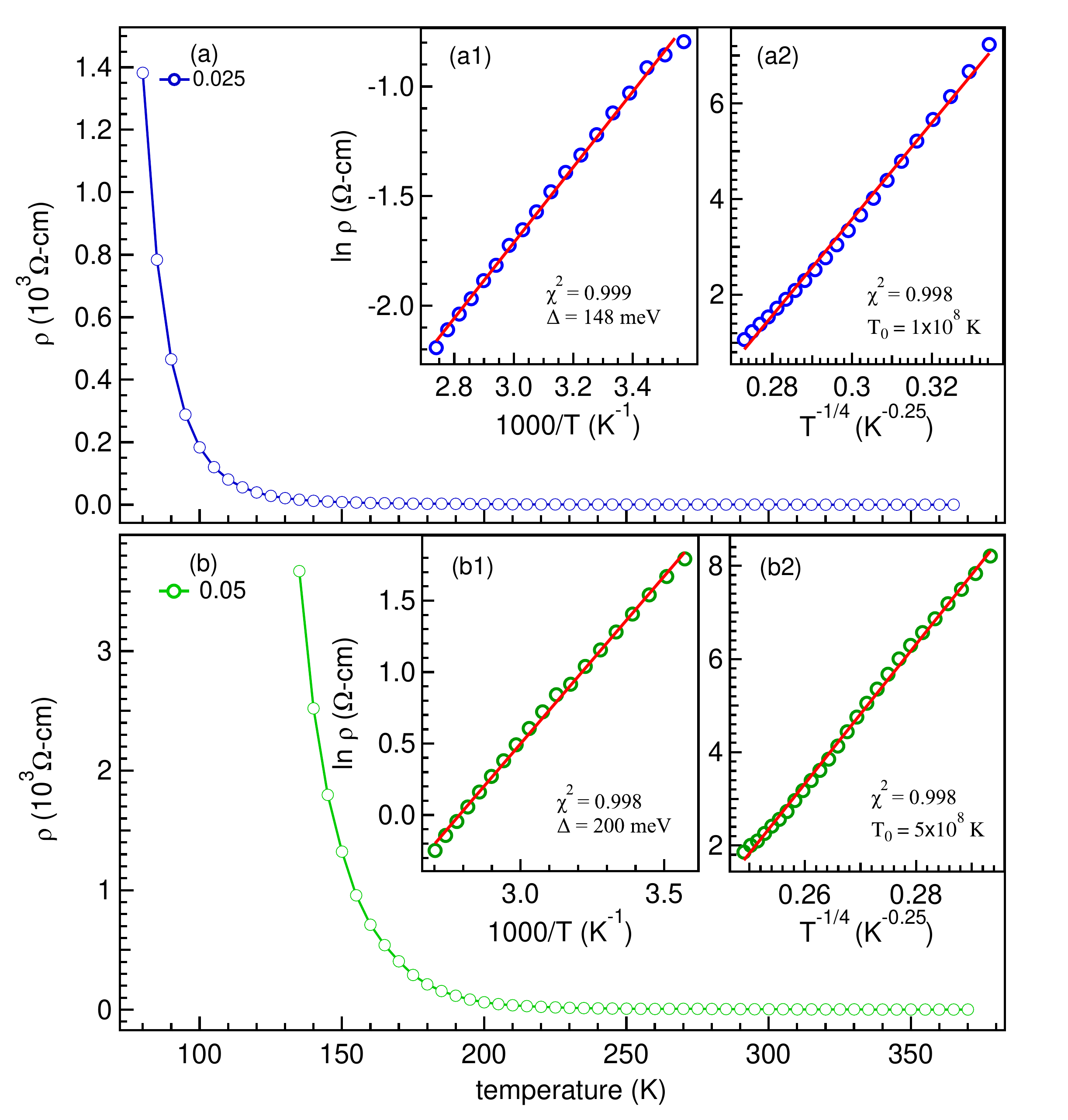}
\caption {Electrical resistance of LaCo$_{1-x}$Nb$_x$O$_3$, $x=$0.025 (a) and 0.05 (b) samples as a function of temperature from 50~K to 370~K. The insets show the plots of ln(R) versus 1000/T and ln($\rho$) versus T$^{-1/4}$, for both the samples. The solid lines through the data are fits using equations~1 (in left panels) and 2 (in right panels).}
\label{fig8:Res}
\end{figure}  

Now we discuss the underlying transport mechanism and measure the temperature dependence of electrical resistivity ($\rho$) of LaCo$_{1-x}$Nb$_x$O$_3$, $x=$0.025 and 0.05 samples, as shown in Figs.~8(a, b). It is clear that both the samples show semiconducting/insulating behavior with decreasing the temperature. More interestingly, the resistivity increase sharply at low temperatures i.e. below about 150~K and 200~K for $x=$0.025 and 0.05 samples, respectively, yielding an insulating ground state, see Figs.~8(a, b). To explain the electrical conduction we use different models and find that no single model can fit the data in the entire temperature range, which suggests at least two types of mechanisms controlling the conduction in the different temperature range. The Arrhenius model describes conduction by simple activation of charge carriers through the band gap between conduction and valance band at high temperatures, as given in the equation~2, 
\begin{equation}
\rho (T) = \rho_0 exp(\Delta/k_BT)
\end{equation}
where $\Delta$ is the activation energy. We plotted ln($\rho$) vs. 1/T in the temperature range from 370 to 280~K [see Figs.~8(a1, b1)] and $\Delta$ values can be obtained from the slope of the curves, which are 148~meV and 200~meV for $x=$ 0.025 and 0.05 samples, respectively. These values of activation energy are significantly larger than the reported 120~meV for $x=$ 0 sample in ref.~\cite{Vinod13} and consistent for insulating transition metal oxides. At lower temperatures the carrier transport is provided by hoping between the localized states. Therefore, we used the Mott's variable range hopping (VRH) model, as described in equation~3, in the lower temperature range close to the transition (from 140 to 85 and 185 to 135~K for $x=$0.025 and 0.05 samples, respectively), see Figs.~8(a2, b2), 
\begin{equation}
\rho (T) = \rho_0 exp(T_0/T)^{1/4}
\end{equation}
where T$_0$ = $18/{k_{B}N(E{_F})\lambda^{3}}$ is the characteristic temperature, and N(E$_F$) is the effective density of states (DOS) near the Fermi level, and $\lambda$ is localization length. Here T$_0$ value is calculated from the slope of ln($\rho$) vs. T$^{-1/4}$ plots, which comes out to be 1$\times$10$^8$~K and 5$\times$10$^8$~K. And, using the T$_0$ values and the $\lambda=$ 2 \AA~ (taken as the average Co--O bond length for LaCoO$_3$), we calculated the density of states [DOS; N(E$_{\rm F}$)] near the Fermi level, which are 2.5$\times10^{20}$ and 5$\times10^{19}$ eV$^{-1}$cm$^{-3}$ for $x=$0.025 and 0.05 samples, respectively. It is clearly evident that the resistivity increases with Nb concentration in the measured temperature range. For example, the $\rho$ is 2~k$\ohm$-cm at 70~K and 5.5~k$\ohm$-cm at 140~K for $x=$0.025 and 0.05 samples, respectively [Figs.~8(a, b)]. The obtained decrease in the DOS values at the Fermi level and increase in the activation energy results in the decrease in conductivity with higher Nb concentration. 

\begin{figure}
\centering
\includegraphics[width=3.55in]{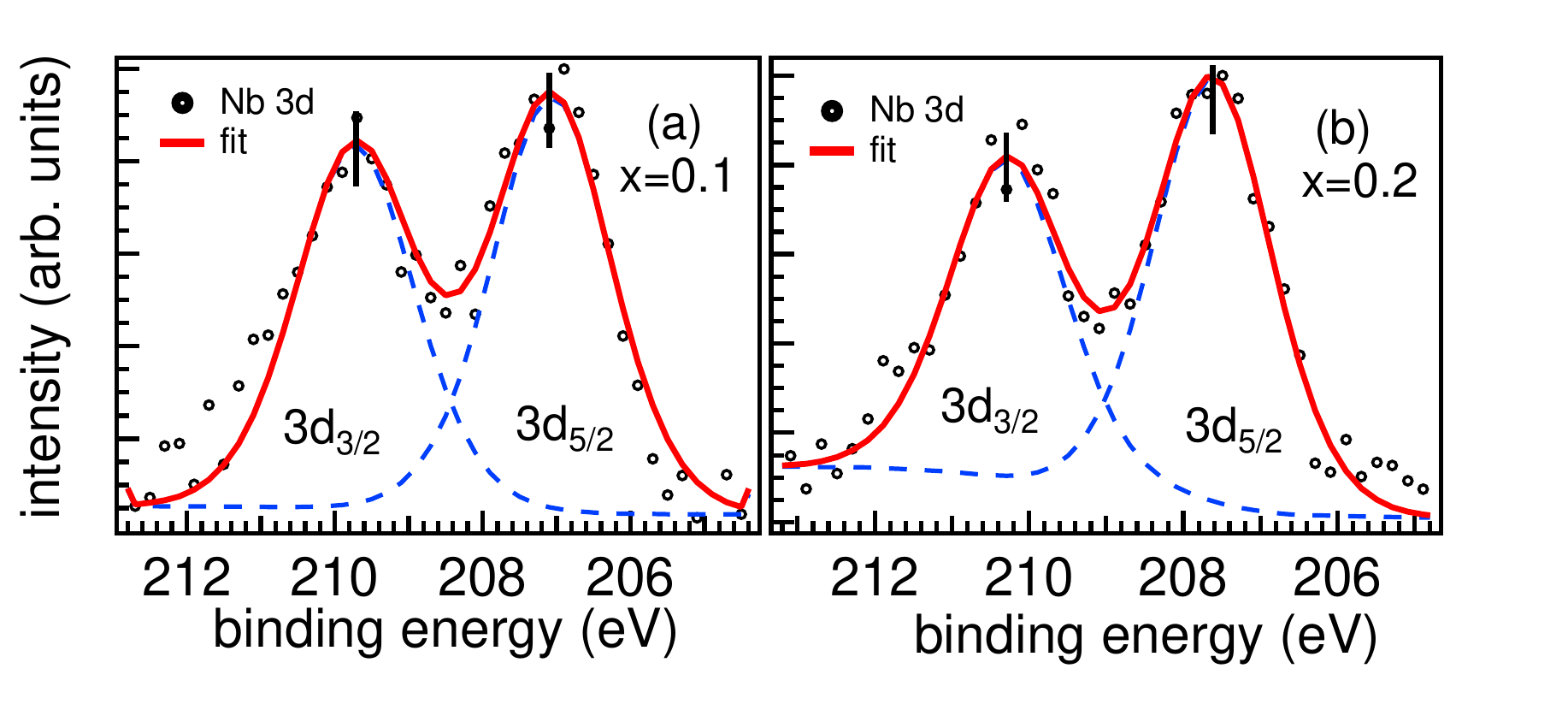}
\caption {The Nb 3$d$ core-level spectra of LaCo$_{1-x}$Nb$_x$O$_3$, $x=$ 0.1 (a), 0.2 (b) samples.}
\label{fig9:Nb3d}
\end{figure}

Further, we use x-ray photoelectron spectroscopy and measured the Co 2$p$ and La 3$d$ core-level spectra (not shown), which are consistent with the previous reports \cite{Ravi17, VasquezPRB96, Chainani92, Saitoh97, LamPRB80, BarmanPRB94}. There are some signature of the presence of Co$^{2+}$ in the Co 2$p$ core-level spectra; however, high-resolution measurements using synchrotron radiation facility are required for detailed analysis. In Figs.~9(a, b), we show the Nb 3$d$ core-level spectra along the fitting with Voigt function after subtracting the inelastic Tougaard background. For $x=$ 0.1 sample, the spin-orbit splitting components Nb 3$d_{5/2}$ and 3$d_{3/2}$ appear at binding energy (BE) 207~eV and 210~eV, respectively. These values confirm that the Nb is predominantly in 5+ state \cite{SasaharaJPCC13}. However,  for $x=$ 0.2 sample we observed the Nb 3$d$ core-levels at about 0.5~eV towards higher BE due to the change in the surrounding chemical environment as higher Nb content converts Co$^{3+}$ ions to Co$^{2+}$ and stabilize in high-spin state.

\section{\noindent ~Conclusions}
In conclusion, the structural, magnetic, transport and electronic properties of LaCo$_{1-x}$Nb$_x$O$_3$ ($x =$ 0--0.2) have been studied in detail using x-ray diffraction, magnetization, DC resistivity and x-ray photoemission spectroscopy. The Rietveld refinements of the room temperature powder x-ray diffraction (XRD) data show structural phase transitions with increasing Nb concentration where coexistence of rhombohedral and orthorhombic is found in $x=$ 0.1--0.15 samples and $x=$ 0.2 sample is present in orthorhombic and monoclinic phases. More interestingly, we observed dramatic changes in the magnetic behavior for Nb concentration as low as 2.5\% where the magnetization increases sharply below $\approx$10~K without bifurcation between ZFC and FC. Furthermore, the spin glass behavior has been observed for $x=$ 0.1 sample in ac susceptibility measurements. The analysis of resistivity data show decrease in the density of state at the Fermi level and increase in the activation energy, which results in strong insulating nature with increasing the Nb concentration. The XPS study of Nb 3$d$ core-levels confirms that the Nb is present mostly in 5+ valence state and shows about 0.5~eV shift towards higher BE side for $x=$ 0.2 sample as the chemical surrounding changes due to more of Co$^{2+}$ ions with Nb substitution. Our results demonstrate that the nonmagnetic Nb$^{5+}$ substitution will convert Co$^{3+}$ ions to Co$^{2+}$ and stabilize in the high-spin state. We find that the spin-state transition and the difference in the ionic radii between Nb$^{5+}$ and Co$^{3+}$/Co$^{2+}$ are crucial in controlling the physical properties of LaCo$_{1-x}$Nb$_x$O$_3$.

\section*{\noindent ~Acknowledgments}

RS acknowledges the MHRD, India for fellowship through IIT Delhi. We thank V. K. Anand, Priyanka, A. K. Pramanik, Mahesh Chandra, Anita Dhaka and Milan Radovic for useful discussions. Authors acknowledge IIT Delhi for providing research facilities: XRD, ÒPPMS EVERCOOL-IIÓ, SQUID, Raman and XPS. We also thank the physics department, IIT Delhi for support. RSD gratefully acknowledges the financial support from SERB-DST through Early Career Research (ECR) Award (project reference no. ECR/2015/000159) and BRNS through ÒDAE Young Scientist Research AwardÓ project sanction No. 34/20/12/2015/BRNS.

\end{document}